\documentclass[%
 reprint,
superscriptaddress,
 amsmath,amssymb,
 aps,
]{revtex4-1}

\usepackage{graphicx}% Include figure files
\usepackage{dcolumn}% Align table columns on decimal point
\usepackage{bm}% bold math
\begin{document}

\preprint{APS/123-QED}

\title{Patterning droplets with durotaxis}% Force line breaks with \\

\author{Robert W. Style}
%\email[]{rob.style@yale.edu}%Lines break automatically or can be forced with \\
\affiliation{%
Yale University, New Haven, CT 06520, USA 
}%

\author{Yonglu Che}%
\affiliation{%
Yale University, New Haven, CT 06520, USA
}

\author{Su Ji Park}%
\affiliation{%
X-ray Imaging Center, Department of Materials Science and Engineering, Pohang University of Science and Technology, San 31, Pohang 790-784, Korea
}

\author{Byung Mook Weon}%
\affiliation{%
School of Advanced Materials Science and Engineering, Sungkyunkwan University, Suwon 440-746, Korea
}

\author{Jung Ho Je}%
\affiliation{%
X-ray Imaging Center, Department of Materials Science and Engineering, Pohang University of Science and Technology, San 31, Pohang 790-784, Korea
}

\author{Callen Hyland}%
\affiliation{%
Yale University, New Haven, CT 06520, USA
}

\author{Guy K. German}%
\affiliation{%
Yale University, New Haven, CT 06520, USA
}

\author{Michael Rooks}%
\affiliation{%
Yale University, New Haven, CT 06520, USA
}

\author{Larry A. Wilen}%
\affiliation{%
Unilever Research and Development, Trumbull, CT 06611, USA
}

\author{John S. Wettlaufer}%
\affiliation{%
Yale University, New Haven, CT 06520, USA 
}
\affiliation{NORDITA, Roslagstullsbacken 23, 10691 Stockholm, Sweden}

\author{Eric R. Dufresne}%
\email[]{eric.dufresne@yale.edu}
\affiliation{%
Yale University, New Haven, CT 06520, USA 
}

%\affiliation{
%\affiliation{
%\affiliation{%

\date{\today}% It is always \today, today,

\begin{abstract}
Numerous cell-types have shown a remarkable ability to detect and move along gradients in stiffness of an underlying substrate -- a process known as durotaxis. The mechanisms underlying durotaxis are still unresolved, but generally believed to involve active sensing and locomotion. Here, we show that simple liquid droplets also undergo durotaxis. By modulating substrate stiffness, we obtain fine control of droplet position on soft, flat substrates. Unlike other control mechanisms, droplet durotaxis works without imposing chemical, thermal, electrical or topographical gradients. This enables a new approach to large-scale droplet patterning and is potentially useful for many applications, such as microfluidics, thermal control and microfabrication. 
\end{abstract}
%\begin{description}
%\end{description}
%\end{abstract}

\pacs{Valid PACS appear here}% PACS, the Physics and Astronomy
\maketitle

\section*{Introduction}
The control of liquids on surfaces is essential for microfluidics  \cite{squi05}, microfabrication \cite{srin01} and coatings \cite{chau92,wong11,soku10}, to name but a few applications. 
%Two primary strategies have emerged to control droplets on surfaces. 
Wetting is typically manipulated by controlling interfacial energies \cite{dege10}. 
Heterogeneous surface chemistries have been exploited to pattern \cite{gau99,sirr00} and transport droplets \cite{chau92,darh05}.
Gradients in temperature  or electric potential can drive droplet motion \cite{darh05,chau92}.
Alternatively, surface topography can  control the spreading of fluids.
For example, isotropically rough surfaces can exhibit superhydrophobicity \cite{cass44,lafu03}, while anisotropic surfaces exhibit anisotropic spreading \cite{cour07} and even directed droplet transport \cite{shas06,lagu11,prak08}. 
Here, we introduce a  method to control droplets on surfaces inspired by the biological phenomenon of durotaxis -- the ability of many eukaryotic cell types to move along gradients in the stiffness of their extracellular matrix \cite{lo00,disc05,tric12,choi12}.
While the current explanation of durotaxis involves active sensing of matrix stiffness and actomyosin-based motility \cite{tric12}, we show here that even simple liquid droplets display durotaxis.
Furthermore, we show that durotaxis can be exploited to achieve large-scale droplet patterning.
A simple theory explains how drops move towards softer parts of a substrate, and quantitatively captures the droplet distribution on patterned surfaces.
Droplet durotaxis is prominent on soft substrates, which are significantly deformed by liquid surface tension \cite{roma10,mora10}.

The spreading of liquid droplets on stiff, flat surfaces is primarily described by the  contact angle.
In equilibrium, a small droplet takes the shape of a spherical cap with uniform contact angle $\theta$ determined by Young's law: $\gamma_{LV} \cos\theta = \gamma_{SV}-\gamma_{SL}$. Here indices $L$, $S$ and $V$ of interfacial energies, $\gamma$, represent liquid, solid and vapor  respectively \cite{dege10}.
Spontaneous droplet motion typically occurs in two main cases.
First, if the actual contact angle of a droplet differs from its equilibrium contact angle, the droplet will be driven to spread/contract until it reaches its equilibrium shape \cite{dege10}.
Second, if there is a difference between the equilibrium contact angle on either side of a droplet, the droplet will be driven towards the more wetting side. This is typically achieved by modifying the interfacial energies across the droplet \cite{darh05}. Here, we show that gradients in substrate stiffness can also drive droplet motion on soft substrates.

\section*{Contact-angle dependence on substrate stiffness}
On soft substrates, the apparent contact angle varies with droplet size and substrate stiffness \cite{styl12,styl12c}.
This breakdown of Young's law occurs because droplet surface tension can significantly deform soft substrates, as shown by the x-ray micrograph in 
Figure 1(a) \cite{shan86,shan87,carr96,extr96,peri08,peri09,jeri11,marc12,lima12,luba12,styl12,styl12c}. 
Droplet surface tension pulls up at the contact line creating a ridge, while the droplet's internal (Laplace) pressure pushes down into the substrate over the contact area, creating a dimple.
Dimple formation leads to a contact-angle change: when the Laplace pressure is sufficiently large, the droplet bulges down into the substrate, taking a lenticular shape and causing it to appear more wetting, as illustrated in Figure 1(b,c).
The apparent contact angle, $\theta$, is defined here as the angle of the liquid-vapor interface at the contact line relative to the undeformed solid surface far from the droplet.
For large droplets, we expect the contact angle to be consistent with Young's law.
 For very small drops, the shape mimics that of drops floating on a liquid with the same interfacial tensions as the soft substrate  -- the contact angle is determined from a Neumann triangle construction \cite{neum94,styl12,styl12c}.
The critical droplet radius where the apparent contact angle starts to deviate from Young's law is $L=\gamma_{LV}/E$, where $L$ is an elastocapillary length and $E$ is the Young's modulus of the substrate \cite{styl12c}. For hard solids, $L \sim$ molecular scales, so contact-angle changes are insignificant. For soft solids, $L$ can be macroscopic; gels with $E \sim$ kPa have $L\sim 10 \mu$m.

\begin{figure}
\centerline{\includegraphics[width=.5\textwidth]{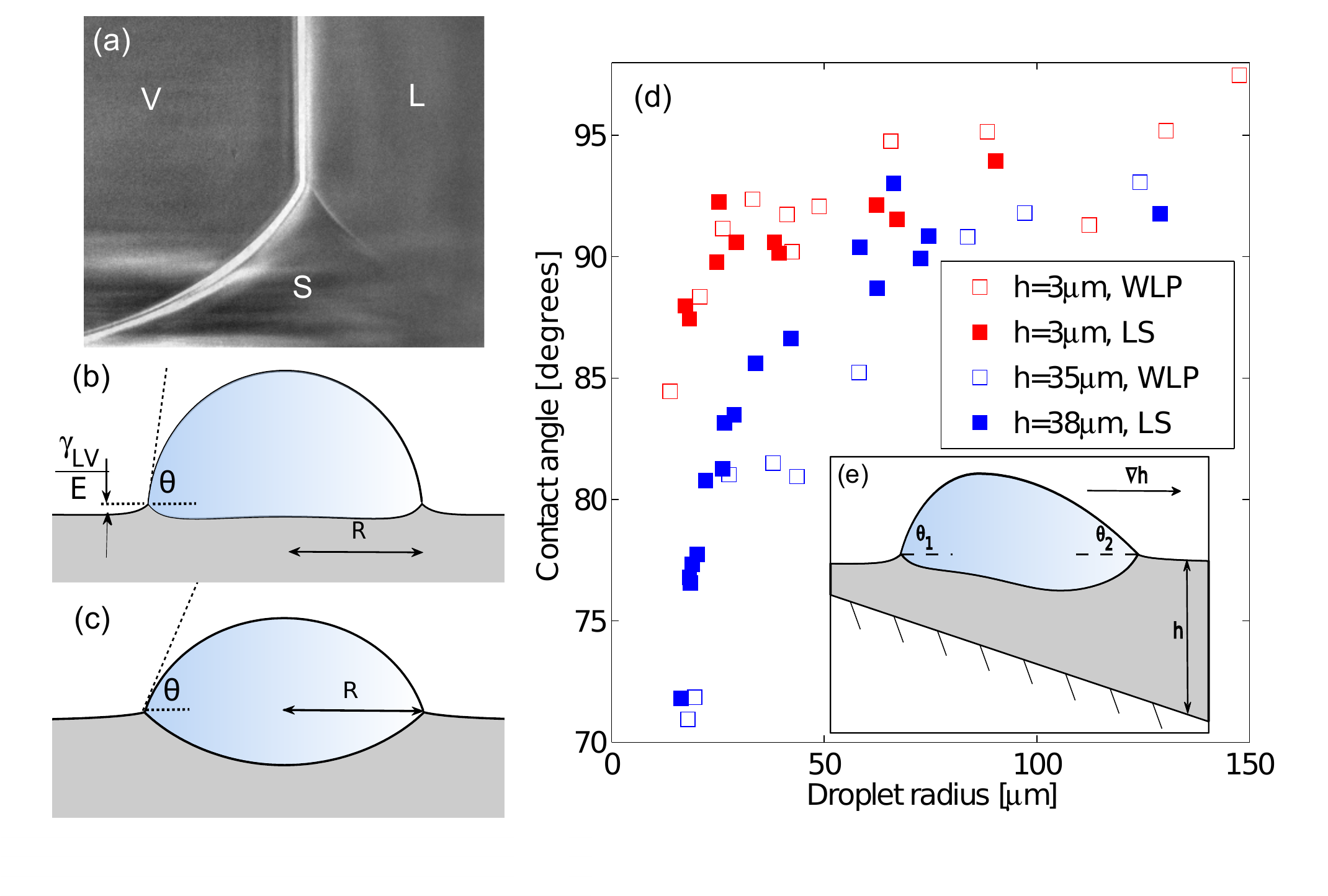}}
\caption{Droplets deform soft substrates, causing Young's law to fail. (a) X-ray image of the contact line of a water droplet on a soft, silicone gel substrate. The ridge is pulled up by the droplet surface tension. $E=3$kPa, and the substrate is $22\mu$m thick. The droplet radius is approximately 1mm. (b) The equilibrium of a sessile droplet on a soft surface with $R\gg\gamma_{LV}/E$ and (c) a soft surface with $R\ll\gamma_{LV}/E$.  (d) Symbols show measured contact angles of glycerol droplets on a silicone gel as a function of droplet radius, $R$.  Data is shown for thin silicone gel layers of $h=3\mu$m (red) and thicker layers $h=35,38\mu$m (blue). Filled/open points were measured by laser scanning (LS)/ white-light optical profilometry (WLP). The large drop contact angle was measured as $95^\circ$ \cite{styl12c}. (e) Schematic profile of a droplet on a soft surface of varying thickness, $h$. }\label{fig:background}
\end{figure}

Two examples of the dependence of the macroscopic contact angle on droplet size are shown in Figure 1(d).
Here, glycerol droplets rest on thin, flat silicone-gel layers (CY52-276A/B, Dow Corning: $E=3$ kPa) spincoated on a stiff glass coverslip. 
In the samples shown, the film thicknesses are $h=3$, 35 and $38\mu$m.
We measured $\theta$ using a laser scan (laser profilometer with white-light probe sensor, Solarius Inc.) and an optical profilometer (NewView 7300, Zygo Corp.). Further details are given in the Materials and Methods.
For droplets larger than 50$\mu$m, $\theta$ approaches the macroscopic value of $95^\circ$, measured for millimetric droplets. 
For droplets smaller than 50$\mu$m, $\theta$ is significantly reduced.
For a given droplet size, the contact angle is smaller on the thicker substrate.
In other words, small droplets appear to wet thick substrates more strongly than thin substrates.  

\section*{Droplet durotaxis}
We hypothesised that  the dependence of  $\theta$ on soft substrate thickness could be exploited to manipulate droplets on chemically-homogeneous, flat surfaces.
The data in  Figure 1(d)  suggest  that a droplet on a soft substrate with non-uniform thickness will have a non-uniform contact angle, as shown in Figure 1(e).
By analogy with droplet motion driven by gradients of interfacial energy, we expect droplets to move spontaneously along gradients in substrate thickness. 
Here, droplet motion is not created by an external force or gradients in interfacial energy, but differences in substrate stiffness.
We use `stiffness' in the sense of a spring constant,  describing how much a surface is displaced by an external force. 
Thus, a substrate's stiffness depends both on its elastic modulus and thickness.
Intriguingly, motion along stiffness gradients has  been observed in living cells.
This phenomenon, called \emph{durotaxis}, can be driven by gradients in the elastic modulus of the substrate \cite{lo00,disc05,tric12} or by gradients in the substrate thickness \cite{choi12}.

\begin{figure}
\centerline{\includegraphics[width=.5\textwidth]{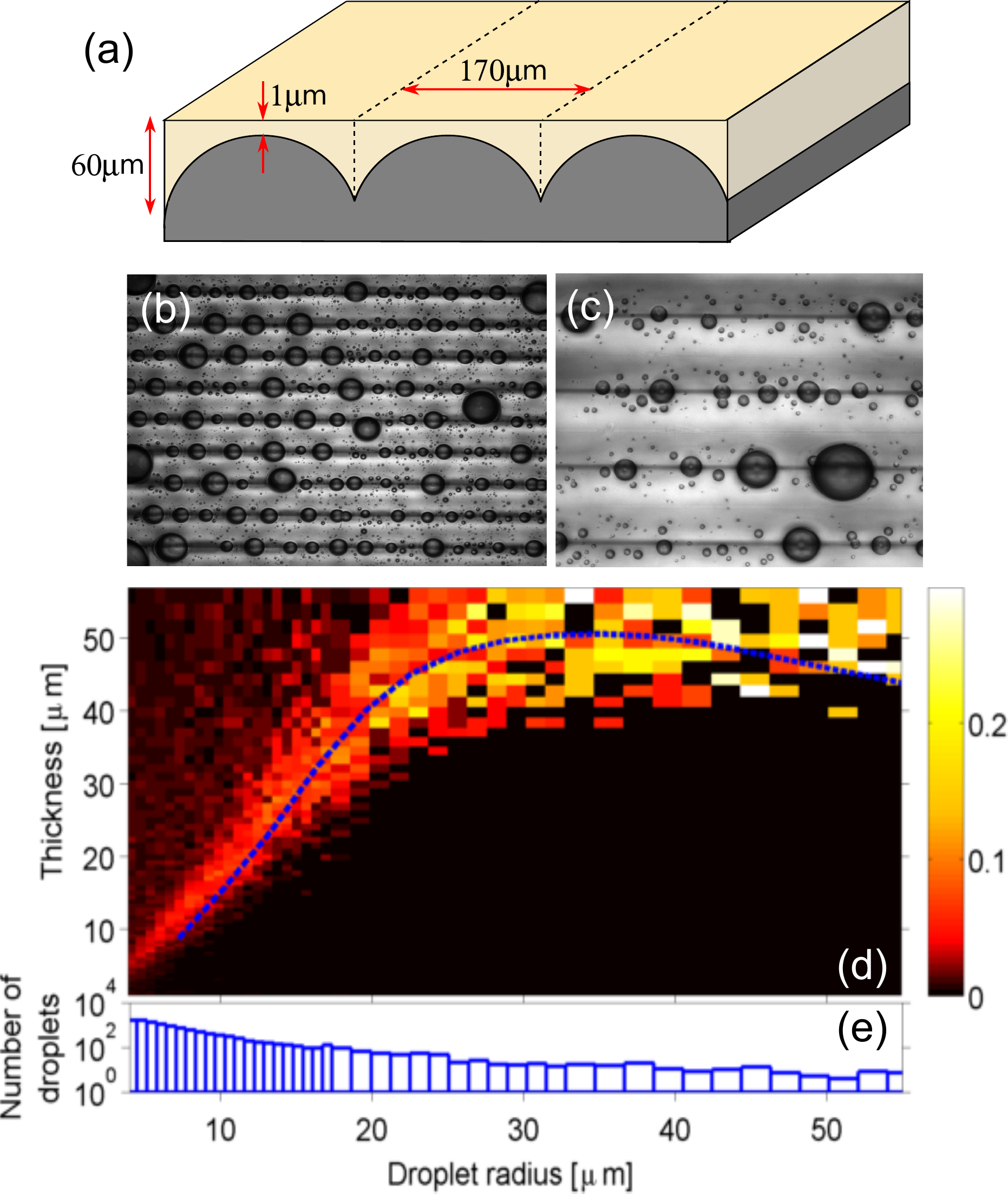}}
  \caption{Droplets move on flat surfaces with stiffness gradients. (a) Schematic diagram of flat chemically-homogeneous substrates with gradients of stiffness.  A flat layer of soft, silicone gel is deposited on a hard, lenticular array creating gradients in the thickness (or stiffness) of the gel. (b,c) Photographs of glycerol droplets after deposition with an atomiser. Dark horizontal bands are located at the thickest regions of the substrate. The spacing between the bands is 170$\mu$m. (b,c) are taken 5 minutes and 5 hours after application respectively. (d) Probability density of final droplet locations from 13300 droplets as a function of substrate thickness and drop radius. The dashed curve is the theoretical prediction for final position of moving droplets. (e) Size distribution of observed droplets.}
  \label{fig:move}
\end{figure}

To test this hypothesis, we observed droplets on soft substrates with flat surfaces and gradients in thickness.
We coated stiff lenticular sheets with silicone gel (Figure 2(a)).
The lenticules are cylindrical caps with radii of 91 $\mu$m spaced periodically with wavelength 170 $\mu$m.
The resulting gel layer had strong thickness gradients but a flat  surface (quantified in the Supporting Information).

We sprayed glycerol droplets onto the surface with an atomiser, and observed them in reflection on a light microscope. Examples are shown in Figures 2(b,c).
The dark lines indicate the deepest part of the substrate.
We found that droplets spontaneously moved from thin regions to thick regions (see {\it SI Movie 1}).
Depending on their size, the droplets continued to move detectably for up to 1 hour, and we left them for 5 hours post-deposition to reach `steady state'.
%Figure \ref{fig:move}(b) shows a photograph of the distribution of glycerol droplets several minutes after deposition by a single burst with an atomiser. 
Figure 2(c) shows a typical steady-state image taken 5 hours post deposition.
We analysed 92 such images, containing 13300 droplets to determine their final positions. Droplet radii and centre positions were detected automatically using a circular-Hough transform (droplets $<20\mu$m in radius), or by identifying points on the droplet perimeter by hand and fitting a circle through these points (larger droplets).
The results are shown in Figure 2(d) as the probability density of droplets ending up on a portion of the substrate of a given thickness. Figure 2(e) shows the number of droplets recorded in each size range.
Large droplets all move to the deepest part of the substrate.
Note that the substrate is $60\mu$m thick at its deepest point.
Droplets below $25\mu$m in radius are most likely to be found where the substrate thickness is roughly 1.5 times their radius; in deeper regions the probability density is roughly uniform and indistinguishable from the uniform coverage by the atomiser.
All droplets are almost perfectly excluded from regions shallower than a drop-size-dependent critical thickness.

Representative droplet trajectories are shown in Figure 3(a), starting within a minute of initial deposition.
This shows droplet velocity as a function of underlying substrate thickness.
Droplets are driven towards thicker regions of the substrate, with speed decreasing as they move.
Small droplets deposited over thick regions of the substrate do not move.

\begin{figure}
\centerline{\includegraphics[width=.5\textwidth]{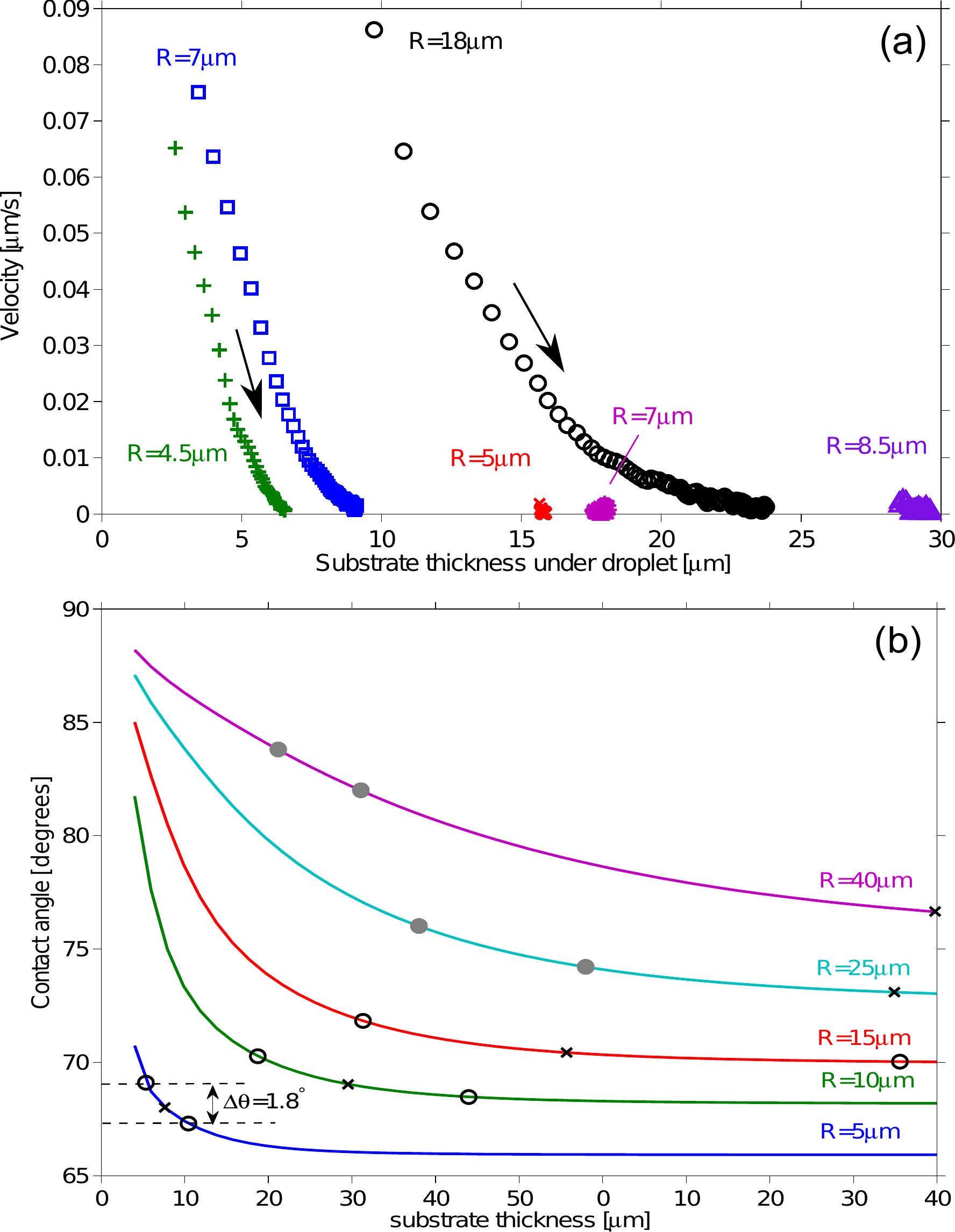}}
  \caption{Droplet motion on stiffness gradients. (a) Speed of droplets moving from thin to thick regions of the gel as a function of gel thickness at droplet centre, $h$.  Trajectories extracted from image sequences such as {\it SI Movie 1}.  Drop speed increased as drop radius increased and substrate thickness decreased.  Droplets came to rest when $1.5 R \approx h$. Droplets starting at locations much thicker than their radii did not move. (b) Solid curves show the theoretical contact angle as a function of substrate thickness for droplets of different sizes. The x-symbols show droplet center and o symbols show droplet front and back at thickness where $\Delta\theta = 1.8^\circ$. The o's are filled when the droplet straddles the trough in the substrate, otherwise empty.}
  \label{fig:prob}
\end{figure}

\section*{Mechanism of Droplet Durotaxis}
A simple theory quantitatively captures the final position of moving droplets.
Extending our analogy with droplet motion driven by gradients of interfacial energy, we expect droplets to move spontaneously when the contact-angle difference across the droplet, $\Delta\theta$, exceeds a critical value, $\Delta\theta_c$ \cite{dege10}. 
Using our previous theory \cite{styl12,styl12c}, we calculated the expected contact angle of a droplet as a function of substrate thickness, as shown by the curves in Figure 3(b) (see Supporting Information).
This was used to estimate $\Delta\theta$ as a function of  droplet size and position, as shown by the gray and black symbols in the figure.
In the shallower parts of the substrate, $\Delta\theta$ can be as much as $15^\circ$, depending on the droplet size. 
As the substrate thickness increases, $\Delta\theta$ decreases.
Thus, the presumed driving force for droplet motion decreases with increasing thickness.
Assuming that the droplet stops when $\Delta\theta=\Delta\theta_c$ we can determine the final position of the droplet as a function of its radius.
Superimposing the theoretical final droplet positions on top of the experimental data in Figure 2(d), we find good agreement between theory and experiments for a critical value $\Delta\theta_c=1.8^\circ$.
This is consistent with contact-angle hysteresis measurements for macroscopic glycerol droplets on flat, silicone-gel-coated surfaces using a contact angle goniometer (VCA Optima, AST Products).
Hysteresis was sufficiently small as to be undetectable within the accuracy of the machine ($\pm 2^\circ$).

In order to confirm this mechanism and rule out other causes of droplet motion, we repeated our experiments replacing the soft silicone gel with a stiff silicone elastomer (Sylgard 184 with 10:1 ratio, Dow Corning: $E \approx $1.8MPa).  For this material,  our theory predicts no significant substrate deformation or contact-angle differences. Indeed, we found no droplet motion, as shown in SI Movie 2.

\section*{Patterning with durotaxis}
The images in Figures 2(b,c) demonstrate the potential of droplet durotaxis for controlled pattern formation.
This is greatly enhanced when droplets are allowed to coarsen, coalesce, and evaporate.
Figure 4(a) shows water droplets on the substrate studied above, where droplets have been condensed from the ambient atmosphere by cooling of the substrate.
 While droplets grow by condensation, long-range forces -- likely elastic in origin -- drive their coalescence ({\it SI Movie 3}).
This growth rapidly drives droplets to the thickest region of the soft substrate.
With the removal of cooling, the remaining small  droplets  rapidly evaporate, leaving the largest droplets precisely localised over the deepest regions of the substrate.
Figure 4(b) shows water droplets condensed onto a flat, silicone layer coated on an arbitrary substrate pattern -- in this case the letter `Y', created by a flat $35\mu$m deep etch into a silicon wafer. The silicone layer is a few microns thick on the wafer surrounding the `Y'. 

\begin{figure}
\centerline{\includegraphics[width=.5\textwidth]{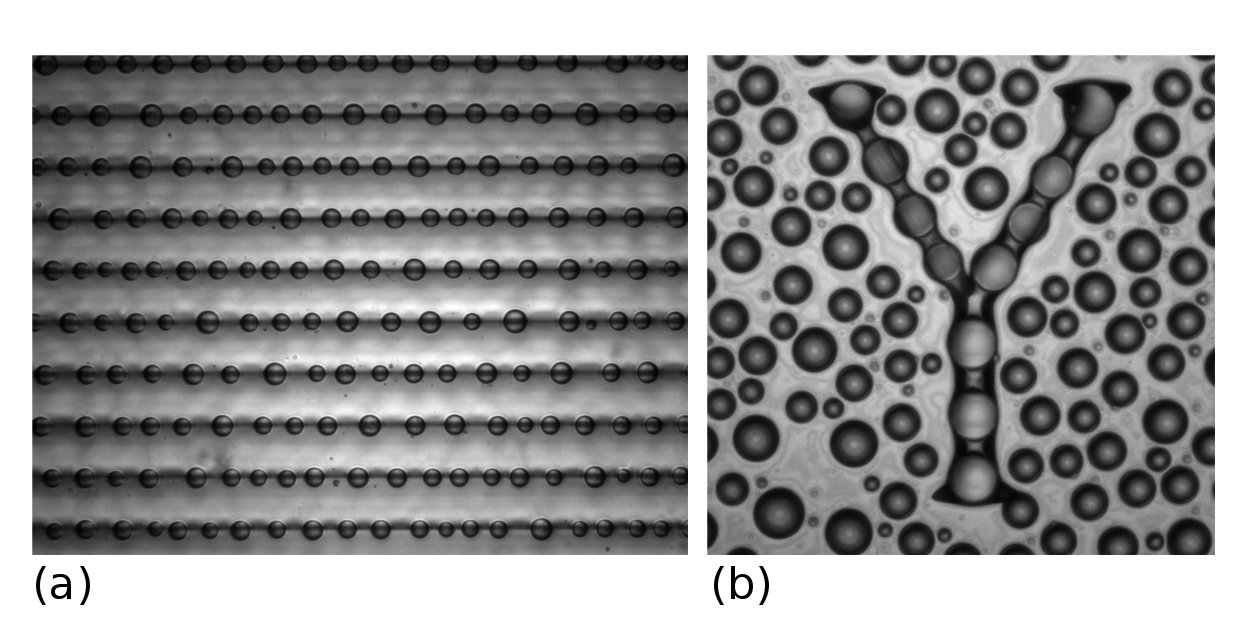}}
  \caption{Patterning droplets with durotaxis. (a) Photograph of water droplets deposited by condensation onto a flat chemically homogeneous surface with varying stiffness.  Dark horizontal bands are located at the thickest regions of the substrate.  The spacing between these lines is 170$\mu$m. The patterning process is shown in the Supporting Information. (b) Droplets deposited by condensation onto soft, flat surface coated a `Y' shape etched into a silicon wafer. Field of view is $620\mu$m wide.}
  \label{fig:patt}
\end{figure}

\section*{Conclusions}
In conclusion, simple liquid droplets undergo durotaxis on stiffness gradients on soft substrates.
This motion can be used to move droplets without chemical, thermal or topographical gradients.
Durotaxis can be used to pattern droplets over large scales, and may also be useful for microfluidics  \cite{darh05}, microfabrication and self-assembly \cite{srin01,wida94}, and condensers \cite{soku10}.
We note that the behaviour of droplets on stiffness gradients is analogous to the collective behaviour of adatoms adsorbed onto the surface of a crystal \cite{vill95}.  Thus droplets  may be a convenient macroscopic analogue for atomic adsorption and redistribution.
%Our results suggest a fresh look at biological durotaxis may be warranted.
Our results also have implications for the biological mechanisms involved in cellular durotaxis.
Even though drops move to softer substrates and cells typically migrate to stiffer substrates, the phenomenon of durotaxis by simple liquid droplets indicates that active stiffness sensing may not be required.

%\section{Digital RNA SNP Analysis} A real-time PCR assay was designed
%to amplify {\it PLAC4} mRNA, with the two SNP alleles being discriminated
%by TaqMan probes. {\it PLAC4} mRNA concentrations were quantified in
%extracted RNA samples followed by dilutions to approximately one target
%template molecule of either type (i.e., either allele) per well.
%Details are given in the {\it SI Materials and Methods}.
%
%\section{Digital RCD Analysis} Extracted DNA was quantified by
%spectrophotometry (NanoDrop Technologies, Wilmington, DE) and diluted to a
%concentration of 
%approximately one target template from either chr21 or ch1 per well.
%\end{materials}
%
%\appendix[Estimating the Spectral Norm of a Matrix]
%In this appendix we describe a method for the estimation of the spectral norm
%of matrix $A$. The method does not require access to the individual
%entries of $A$; it requires only applications of $A$ and $A$* to vectors.
%It is a version of the classical power method. Its probabilistic
%analysis summarized below was introduced fairly recently in refs. 13
%and 14. This appendix is included here for completeness.
%
%
%\appendix
%This is an example of an appendix without a title.
{\footnotesize
\begin{acknowledgments}
RWS is funded by the Yale University Bateman Interdepartmental Postdoctoral Fellowship. JSW thanks the Swedish Research Council for support. Support for instrumentation was provided by NSF (DBI-0619674). This research was also supported by the Yale Institute for Nanoscience and Quantum Engineering, and by the Creative Research Initiatives (Functional X-ray Imaging) of MEST/NRF. Use of the Advanced Photon Source, an Office of Science User Facility operated for the U.S. Department of Energy (DOE) Office of Science by Argonne National Laboratory, was supported by the U.S. DOE under Contract No. DE-AC02-06CH11357. We thank Zygo Corp. for the use of their optical profilometer and help with measuring contact angles.  We thank Valerie Horsley and Anand Jagota for helpful conversations.
\end{acknowledgments}}

{\footnotesize \section*{Materials and Methods}
{\bf Substrate fabrication} We created our substrates by coating a flat layer of silicone on a lenticular, polyester-resin substrate (Lenstar Plus - Thin, Pacur). For the soft substrates, we used a silicone gel (CY52-276A/B, Dow Corning), while for the harder substrates we used a silicone elastomer (Sylgard 184). The fabrication process is shown in the Supporting Information. We spincoated a 2(w/v)\% solution of polystyrene in toluene on a glass slide to form a flat coating of polystyrene. The silicone was premixed and degassed in a vacuum chamber and then a small drop was sandwiched between the polystyrene film and the lenticular substrate. A flat weight ($\sim$200g) was placed on the sample to compress the gel layer. After curing overnight, the weight was removed and the coated substrate was gently peeled off the polystyrene.

{\bf X-ray imaging of the contact line} A flat, soft substrate was made by spincoating a $22\mu$m thick layer of the silicone gel on a glass slide. A droplet of pure, de-ionized, water (Millipore, 18M$\Omega$cm at 25$^\circ$C) was placed on the substrate and allowed to equilibrate for several minutes. The contact line was then visualised using x-ray imaging performed using the transmission x-ray microscopy at the 32-ID-C beamline in the Advanced Photon Source (APS) of the Argonne National Laboratory. High-resolution, 50nm/pixel micrographs were recorded using bright light at a photon energy of 8 keV generated by the APS \cite{chu08}. As we used a short exposure time of less than 0.2s per snapshot, x-ray effects -- such as polymer degradation -- should be negligible \cite{weon09}.  A Zernike phase-ring provided sufficient phase contrast to clearly visualise the droplet interfaces, so no contrast agent was required  \cite{weon11,weon12}.

{\bf Contact angles} We measured droplet contact angles using two separate techniques: First with an optical profilometer (NewView 7300, Zygo Corp.), and second with a laser scan device (laser profilometer with white-light probe sensor, Solarius Inc.). First, we measured the height from the undeformed surface $h_d$ of droplets on a range of substrate thicknesses. Second, we used an image of the droplet to measure its apparent radius, $R$. When $\theta>90^\circ$, $R$ represents the actual radius of the droplet. When $\theta<90^\circ$, $R$ represents the footprint radius of the droplet. 
We calculated the contact angle $\theta$ by assuming that the droplet surface is a spherical cap of height $h_d$ and radius $R$: if $R>h_d$, $(1-\cos\theta)/\sin\theta=h_d/R$, while if $R<h_d$, $\theta=\sin^{-1}((h_d-R)/R)+90^\circ$. On a rigid substrate with no wetting ridge, this technique will give the correct contact angle. However on soft substrates, this value systematically overestimates $\theta$ by an amount on the order the ratio of the ridge height divided by the droplet radius. This systematic error is $\approx 5^\circ$ for the smallest droplets and decreases with increasing radius \cite{styl12c}.}

\end{document}